\documentclass{aastex6}

\usepackage{graphicx}
\usepackage{dcolumn}
\usepackage{bm}

\newcommand{\be}{\begin{equation}}
\newcommand{\ee}{\end{equation}}
\newcommand{\ba}{\begin{eqnarray}}
\newcommand{\ea}{\end{eqnarray}}
\newcommand{\ban}{\begin{eqnarray*}}
\newcommand{\ean}{\end{eqnarray*}}

\usepackage{graphicx}
\usepackage{dcolumn}
\usepackage{bm}
\usepackage{color}
\begin{document}


\preprint{RUP-16-25}
\preprint{KEK-TH-1929}
\preprint{KEK-Cosmo-197}
\preprint{OCU-PHYS-454}
\preprint{AP-GR-134}

\title{Primordial black hole formation 
in the matter-dominated phase of the Universe}




\author{Tomohiro Harada\altaffilmark{1}}
\author{Chul-Moon Yoo\altaffilmark{2}} 
\author{Kazunori Kohri\altaffilmark{3,4}} 
\author{Ken-ichi Nakao\altaffilmark{5}}
\and 
\author{Sanjay Jhingan\altaffilmark{6,7}}%
%
%
\affil{}
\email{harada@rikkyo.ac.jp}
\altaffiltext{1}{Department of Physics, Rikkyo University, Toshima,
Tokyo 171-8501, Japan}
\altaffiltext{2}{Gravity and Particle Cosmology Group,
Division of Particle and Astrophysical Science,
Graduate School of Science, Nagoya University, Nagoya 464-8602, Japan}
\altaffiltext{3}{Institute of Particle and Nuclear Studies, KEK,
1-1 Oho, Tsukuba, Ibaraki 305-0801, Japan}
\altaffiltext{4}{The Graduate University for Advanced Studies (SOKENDAI),
1-1 Oho, Tsukuba, Ibaraki 305-0801, Japan}
\altaffiltext{5}{Department of Mathematics and Physics,
Graduate School of Science, Osaka City University,
3-3-138 Sugimoto, Sumiyoshi, Osaka 558-8585, Japan}
\altaffiltext{6}{iCLA, Yamanashi Gakuin University, 
		2-4-5, Sakaori, Kofu-City, Yamanashi,  400-8575, Japan}
\altaffiltext{7}{Centre for Theoretical Physics, Jamia Millia Islamia,
New Delhi 110025, India}
\shortauthors{Harada, T., et al.}
\shorttitle{Primordial black hole formation in the matter-dominated
phase of the Universe}

\begin{abstract}

We investigate primordial black hole formation in the matter-dominated
 phase of the Universe, where nonspherical effects in gravitational
 collapse play a crucial role. This is in contrast to the 
 black hole formation in a radiation-dominated era.
We apply the Zel'dovich approximation, 
Thorne's hoop conjecture, and 
Doroshkevich's probability distribution
and subsequently derive the production probability $\beta_{0}$ of primordial black holes. 
The numerical result obtained is 
applicable even if the density fluctuation $\sigma$ 
at horizon entry is of the order of unity.
For $\sigma\ll 1$, we find 
a semi-analytic formula $\beta_{0}\simeq 0.05556 \sigma^{5}$, which is 
comparable with the Khlopov-Polnarev formula.
We find that the production probability in the matter-dominated era
is much larger than that in the radiation-dominated era 
for $\sigma\alt 0.05$, while they 
are comparable with each other for $\sigma\agt 0.05$.
We also discuss how $\sigma$
can be written in terms of primordial 
curvature perturbations.

\end{abstract}




\section{Introduction}
Primordial black holes 
are becoming a very important area of study
at the intersection of cosmology, astrophysics, 
high-energy physics, and gravitation.
See \citet{Carr:2003bj} and \citet{Khlopov:2008qy} for recent reviews.
The abundance of primordial black holes is 
severely constrained observationally
\citep{Carr:1975qj,Carr:2009jm,Carr:2016hva} and this fact has 
rich implications to the early Universe and other relevant
fields of physics. Furthermore, 
recently, LIGO 
has reported gravitational wave observation 
GW150914 \citep{Abbott:2016blz} and   
it has been argued \citep{Sasaki:2016jop} that 
a binary system of primordial black holes can be a source of 
gravitational waves of this event.
The theoretical prediction of the abundance of primordial black holes
based on the physical theory of black hole formation 
is a key issue from the theoretical side.

Khlopov and Polnarev \citep{Khlopov:1980mg,Polnarev:1982}
pioneered primordial black hole formation in the 
matter-dominated era of the Universe.
They argued that if stable superheavy particles predicted in the 
grand unified theories dominate the Universe, the pressure of the 
matter field can be effectively neglected and the production of 
primordial black holes is significantly enhanced.
More recently, \citet{Alabidi:2009bk} and
\citet{Alabidi:2012ex,Alabidi:2013wtp} showed that in 
the so-called hill-top type inflation scenario, 
density perturbations of large amplitude can arise on small scales and 
lead to an enhanced formation of primordial black holes 
in an effectively matter-dominated phase of 
the Universe before the reheating phase.

The formation of primordial black holes had been conventionally 
studied in the radiation-dominated era until recently. 
In this case, the threshold $\tilde{\delta}_{c}$ of the 
amplitude of the density perturbation in the comoving slicing 
at horizon entry 
is determined by the Jeans criterion. 
The production probability $\beta_{0}$ of primordial black holes
is given by $\beta_{0}\sim
\sqrt{2/\pi}(\sigma/\tilde{\delta}_{c})
\exp(-\tilde{\delta}_{c}^{2}/2\sigma^{2})$,
where $\sigma$ is the standard deviation of the 
density perturbations at the relevant mass scale at horizon entry. 
The results of numerical relativity simulations in spherical symmetry 
give the threshold 
$\tilde{\delta}_{c}\simeq
0.42-0.50$~\citep{Shibata:1999zs,Musco:2004ak,Polnarev:2006aa,Musco:2012au,Harada:2015yda}. 
Since the threshold value is of the order of unity, 
gravitational instability leads to the formation of a black hole
shortly after the horizon entry. 

For the equation of state $p=w\rho c^{2}$, 
the analytic formula for the threshold gives 
$\tilde{\delta}_{c}\simeq [3(1+w)/(5+3w)]\sin^{2}[\pi\sqrt{w}/(1+3w)]$~\citep{Harada:2013epa}
showing a good agreement with the numerical results for 
$0.01\le w\le 0.6$~\citep{Musco:2012au}.
In the limit of $w\to 0$, we have $\tilde{\delta}_{c}\to 0$, i.e., 
the region which is only slightly overdense would 
necessarily collapse to a black hole. 
This argument clearly overestimates $\beta_{0}$ because 
it neglects nonspherical effects.
For a density perturbation of small amplitude to collapse to a black hole, 
it must shrink to a radius much smaller than that at the 
maximum expansion so that a deviation from spherical symmetry 
can significantly grow. This instability 
generally leads to a ``pancake'' collapse 
\citep{Lin:1965,Zeldovich:1969sb}. 
\citet{Khlopov:1980mg} and \citet{Polnarev:1982}
discussed that a nonspherical effect significantly suppresses primordial 
black hole formation and obtained a compact analytic formula
for $\beta_{0}$
under the assumption of the small density perturbations.
Nonspherical effects on primordial black holes have also recently 
been discussed by \citet{Harada:2015ewt} and \citet{Kuhnel:2016exn}.

The purpose of the current paper is to estimate $\beta_{0}$
even for a large fluctuation of perturbations and a large deviation from 
spherical symmetry based on a physical argument
and also to reproduce the formula by Khlopov and Polnarev in some 
sense by adopting the approximation of small fluctuation.
We apply the Zel'dovich approximation~\citep{Zeldovich:1969sb}, 
which is a well-established analytic approximation
to describe the nonlinear evolution of density perturbations.
See \citet{White:2014gfa} and references therein for its 
validity, application, and limitation.
See also \citet{Russ:1995eu} for its
general relativistic generalization.
We adopt the hoop conjecture proposed by 
\citet{Thorne:1972ji}
for the formation of a black hole 
horizon, which has not yet been proved for general situations but shown 
to hold even for highly distorted horizons~\citep{Yoshino:2007yb}.
See also \citet{Malec:2015oza} for a recent proof for a special case.
However, see~\citet{East:2016anr} 
for its possible violation in the presence of a negative energy density.
As for the probability distribution of nonspherical perturbations,
we adopt Doroshkevich's one~\citep{Doroshkevich:1970}, 
which was derived under a least number of natural assumptions.

This paper is organized as follows. In Sec. II, we apply 
the Zel'dovich approximation to the nonlinear evolution of the density 
perturbations and obtain the criterion of the black hole 
formation based on Thorne's hoop conjecture.
In Sec. III, we introduce the probability distribution for 
nonspherical perturbations by Doroshkevich, 
derive an integral expression for the production probability 
of primordial black holes
without assuming the small fluctuation approximation. 
Moreover, we obtain a semi-analytic formula under 
the small fluctuation approximation.
In Sec. IV we discuss our results followed by conclusions in Sec. V.
We keep both the gravitational constant $G$ and the speed of light $c$
throughout this paper.

\section{Nonspherical collapse of the density perturbations}
\subsection{Zel'dovich approximation}
We begin with the Zel'dovich approximation~\citep{Zeldovich:1969sb}:
\begin{equation}
 r_{i}=a(t)q_{i}+b(t)p_{i}(q_{j}),
\label{eq:Zeldovich_approximation}
\end{equation}
where $a(t)$ ($>0$), $q_{i}$, and $p_{i}(q_{j})$ ($i=1,2,3$) 
are the scale factor, Lagrangian coordinates, and deviation vector,
respectively. The function
$b(t)$ denotes a linearly growing mode 
in the matter-dominated phase of the Universe
in the framework of the Newtonian cosmology. 
The scale factor $a(t)$ satisfies the
Friedmann equation for a spatially flat universe
\begin{equation}
 H^{2}=\frac{8\pi}{3} G \bar{\rho},
\label{eq:Friedmann}
\end{equation} 
where $H:={\dot{a}}/{a}$
and $\bar{\rho}=\bar{\rho}(t)$ is the density of the Friedmann universe. 
The conservation law implies 
$\bar{\rho}\propto 1/a^{3}$.
Although $b(t)$ is a linearly growing mode, the Zel'dovich approximation 
implies the extrapolation of Eq.~(\ref{eq:Zeldovich_approximation}) 
beyond the linear regime to the nonlinear regime up until a caustic occurs at $q_{i}$.

We can calculate the deformation tensor $D_{ik}$ such that
\begin{equation}
 D_{ik}:=\frac{\partial r_{i}}{\partial
  q_{k}}=a(t)\delta_{ik}+b(t)\frac{\partial p_{i}}{\partial q_{k}}.
\end{equation}
The matrix ${\partial p_{i}}/{\partial q_{k}}$ defines a set of 
fundamental axes and we can choose $q_{i}$ so that
\begin{equation}
 \frac{\partial p_{i}}{\partial q_{k}}=\mbox{diag}(-\alpha, -\beta,-\gamma),
\end{equation}
where $\alpha$, $\beta$, and $\gamma$ are functions of $q_{i}$ and, hence, 
\begin{equation}
 D_{ik}=\mbox{diag} (a-\alpha b, a-\beta b, a-\gamma b).
\end{equation}
The mass contained within the Lagrangian volume is conserved,
i.e., 
\begin{equation}
 dm=\rho d^{3}r=\bar{\rho}a^{3}d^{3}q
\quad 
\mbox{or} 
\quad 
m=\int \rho d^{3}r=\bar{\rho}a^{3}\int d^{3}q.
\end{equation}
Therefore, the density $\rho$ is calculated through 
the determinant of $D_{ik}$ so that 
\begin{equation}
 \rho=\frac{a^{3}}{(a-\alpha b)(a-\beta b)(a-\gamma b)}\bar{\rho}.
\end{equation}

For convenience, we define a density perturbation 
in the linear regime 
\begin{equation}
 \delta_{L}:=\left(\frac{\rho-\bar{\rho}}{\bar{\rho}}\right)_{L}=(\alpha+\beta+\gamma)\frac{b}{a}.
\label{eq:delta_L}
\end{equation}
Note that if we take $b>0$, we find that 
$\delta_{L}>0$ if and only if $\alpha+\beta+\gamma>0$.
Since the density perturbation in the linear regime grows as the scale
factor, i.e., $\delta_{L} \propto a $, we find 
\begin{equation}
 b\propto a^{2}.
\label{eq:b_propto_a^2}
\end{equation} 

\subsection{Pancake collapse}
Due to the continuity of $\alpha$, $\beta$, and $\gamma$, we can locally
take the coordinates $q_{i}$ such that
\begin{eqnarray}
 \left\{\begin{array}{c}
  r_{1}=(a-\alpha b)q_{1}\\
  r_{2}=(a-\beta b)q_{2}\\
  r_{3}=(a-\gamma b)q_{3}
	\end{array}
\right. .
\end{eqnarray}
We assume that $\alpha$, $\beta$, and $\gamma$ 
are constant over the scale in which we are interested.
We also assume $\infty>\alpha\ge \beta\ge \gamma>-\infty$ without loss
of generality. We fix the scale of the Lagrangian coordinate radius
and, hence, consider a ball of radius $q$,
which gives the comoving scale 
of the perturbation. We assume that the perturbation will 
collapse at least along one of the three axes so that $\alpha>0$.
We do not assume that a deviation from spherical symmetry is small.

We define three important moments, the horizon entry time $t=t_{i}$, 
maximum expansion time $t=t_{f}$, and 
collapse time $t=t_{c}$.

At the horizon entry $t=t_{i}$, the unperturbed physical 
radius of the mass is equal to the Hubble radius so that 
\begin{equation}
 a(t_{i})q=cH^{-1}(t_{i}).
\label{eq:horizon_entry}
\end{equation} 
Using Eqs.~(\ref{eq:Friedmann}) and (\ref{eq:horizon_entry}), 
we can calculate the mass contained within the radius $q$ to
\begin{equation}
 m=\frac{4\pi}{3}\bar{\rho}(t_{i})a^{3}(t_{i})q^{3}=\frac{c^{3}}{2G H(t_{i})}
\end{equation}
and, hence, from Eq.~(\ref{eq:horizon_entry}), we find  
\begin{equation}
 a(t_{i})q=r_{g},
\end{equation} 
where $r_{g}:={2Gm}/{c^{2}}$ is the gravitational radius of 
the mass $m$.
From Eq.~(\ref{eq:delta_L}), we can get the following relation between
$b(t_{i})$ and $\delta_{L}(t_{i})$:
\begin{equation}
\delta_{L}(t_{i})=(\alpha+\beta+\gamma)\frac{b}{a}(t_{i}).
\label{eq:delta_L_expression}
\end{equation}

At the maximum expansion $t=t_{f}$, the mass
is about to shrink along the $r_{1}$ axis. 
Since $\dot{r}_{1}(t_{f})=0$ or $\dot{a}(t_{f})=\alpha \dot{b}(t_{f})$, we find  
\begin{equation}
\frac{b}{a}(t_{f})=\frac{1}{2\alpha},
\label{eq:t_f}
\end{equation}
where we have used Eq.~(\ref{eq:b_propto_a^2}). The axis 
$r_{f}:=r_{1}(t_{f})$ can be calculated to 
\begin{equation}
r_{f}=[a(t_{f})-\alpha b(t_{f})]q=\frac{1}{2}a(t_{f})q.
\end{equation}
We denote the ratio of $r_{g}$ to $r_{f}$ with $\chi$, which is given by 
\begin{equation}
 \chi:=\frac{r_{g}}{r_{f}}=2\frac{a(t_{i})}{a(t_{f})}
=4\alpha\frac{b}{a}(t_{i})=
\frac{4\alpha}{\alpha+\beta+\gamma}\delta_{L}(t_{i}).
\label{eq:r_g_to_r_f}
\end{equation}
This clearly shows that a perturbation of small amplitude 
must considerably shrink until it collapses 
to a black hole.
It is also instructive to see the ratio of the radius of the 
mass to the Hubble radius at the time of maximum expansion as follows:
\begin{equation}
 \frac{r_{f}}{cH^{-1}(t_{f})}=\left(\frac{1}{2}\right)^{3/2}\chi^{1/2}
=\left(\frac{1}{2}\right)^{3/2}\left[\frac{4\alpha}{\alpha+\beta+\gamma}\delta_{L}(t_{i})\right]^{1/2},
\end{equation}
where we have used $a\propto t^{2/3}$.
This means that if $\chi\ll 1$ or 
$\delta_{L}(t_{i})\ll 1$, the density perturbation 
grows to the order of unity only after the radius of the mass 
becomes much smaller than the Hubble radius. This justifies the use of 
the Zel'dovich approximation 
based on the Newtonian gravity in the present setting.

At the collapse $t=t_{c}$, $r_{1}(t_{c})=0$ so that 
\begin{equation}
\frac{b}{a}(t_{c})=\frac{1}{\alpha}.
\label{eq:t_c}
\end{equation}
Then, the mass becomes a ``pancake'' or a two-dimensional ellipse 
with the semi-minor and semi-major axes given by 
\begin{equation}
 r_{2}(t_{c})=[a(t_{c})-\beta b(t_{c})]q \quad {\rm and} \quad 
 r_{3}(t_{c})=[a(t_{c})-\gamma b(t_{c})]q,
\label{eq:r_2_r_3_t_c}
\end{equation}
respectively. From Eqs.~(\ref{eq:b_propto_a^2}), 
(\ref{eq:t_f}), and (\ref{eq:t_c}), we find 
$a(t_{c})q=2a(t_{f})q=4r_{f}$. Using Eqs.~(\ref{eq:t_c}) and 
(\ref{eq:r_2_r_3_t_c}), we find
\begin{equation}
 r_{2}(t_{c})=4\left(1-\frac{\beta}{\alpha}\right)r_{f} 
\quad {\rm and}\quad  
 r_{3}(t_{c})=4\left(1-\frac{\gamma}{\alpha}\right)r_{f}.
\end{equation}

\subsection{Black hole formation criterion}
The hoop conjecture 
\citep{Thorne:1972ji,Misner:1974qy} 
states that black holes with horizons form when 
and only when a mass $M$ gets compactified 
into a region whose circumference in every direction
is approximately smaller than $4\pi G M/c^{2}$.
The hoop ${\cal C}$ of a region is defined as 
the maximum of its circumferences in all directions.
For the pancake, it is given by 
the circumference of the ellipse.
Since the eccentricity of the pancake is given by 
\begin{equation}
 e^{2}=1-\left(\frac{r_{2}(t_{c})}{r_{3}(t_{c})}\right)^{2}=1-\left(\frac{\alpha-\beta}{\alpha-\gamma}\right)^{2},
\label{eq:eccentricity}
\end{equation}
the hoop is calculated to 
\begin{equation}
{\cal
 C}=16\left(1-\frac{\gamma}{\alpha}\right)E\left(\sqrt{1-\left(\frac{\alpha-\beta}{\alpha-\gamma}\right)^{2}}\right)r_{f},
\label{eq:hoop}
\end{equation}
where $E(e)$ is the complete elliptic integral of the second kind.
Note that $E(e)$ is a monotonically decreasing function of $e\in [0,1]$,
where $E(0)=\pi/2$ in the circular limit and $E(1)=1$ in the eccentric limit.

According to the hoop conjecture, the condition for black hole formation
is given by ${\cal C}\lesssim 2\pi r_{g} $.
Thus, we find the criterion  
\begin{equation}
s(\alpha,\beta,\gamma)\lesssim \chi,
\end{equation}
where 
\begin{equation}
 s(\alpha,\beta,\gamma):=\frac{8}{\pi}\left(1-\frac{\gamma}{\alpha}\right)E\left(\sqrt{1-\left(\frac{\alpha-\beta}{\alpha-\gamma}\right)^{2}}\right).
\end{equation}
Equivalently, 
we can rewrite the criterion in the following form:
\begin{equation}
 h(\alpha,\beta,\gamma)\lesssim 1, 
\label{eq:criterion_3}
\end{equation}
where
we define $h(\alpha,\beta,\gamma):={\cal C}/(2\pi r_{g})$
according to \citet{Yoshino:2007yb}. 
Using Eq.~(\ref{eq:r_g_to_r_f}), we can calculate the ratio to
\begin{equation}
 h(\alpha,\beta,\gamma)=
\frac{2}{\pi}\frac{\alpha-\gamma}{\alpha^{2}}
E\left(\sqrt{1-\left(\frac{\alpha-\beta}{\alpha-\gamma}\right)^{2}}\right),
\label{eq:s_3}
\end{equation}
where we have fixed the normalization of $b$ so that 
\begin{equation}
 \frac{b}{a}(t)=\frac{a(t)}{a(t_{i})}.
\label{eq:normalisation_b}
\end{equation}
The set of Eqs.~(\ref{eq:criterion_3}) and (\ref{eq:s_3})
is written only in terms of $\alpha$, $\beta$, and $\gamma$
and, hence, is the most suitable expressions describing 
the present situation.

If the above criterion is not
satisfied, a sheet-like caustic occurs at $t=t_{c}$ and 
matter particles should cross each other.
These pancakes will then undergo violent relaxation 
by bouncing several times and eventually get virialized with 
a large velocity dispersion.
The radius of such a virialized object is approximately 
half the radius of maximum expansion $r_{f}$. 
The object may later shrink and even collapse to 
a strongly bound object 
by radiating its energy through some form of radiation.
This process happens in a time scale much larger 
than the Hubble time.

\section{
Production probability of primordial black holes}
\label{sec:probability}
\subsection{Doroshkevich's probability distribution}
The probability distribution of $\alpha$, $\beta$, and $\gamma$ is 
given by \citet{Doroshkevich:1970} as 
\begin{eqnarray}
&& w(\alpha,\beta,\gamma)d\alpha d\beta d\gamma \nonumber \\
&=& -\frac{27}{8\sqrt{5}\pi \sigma_{3}^{6}}\exp
\left[-\frac{3}{5\sigma_{3}^{2}}\left\{(\alpha^{2}+\beta^{2}+\gamma^{2})
-\frac{1}{2}(\alpha\beta+\beta\gamma+\gamma\alpha)\right\}
\right] \nonumber \\
&& \cdot (\alpha-\beta)(\beta-\gamma)(\gamma-\alpha)d\alpha d\beta d\gamma,
\label{eq:Doroshkevich_distribution}
\end{eqnarray}
where $\infty>\alpha\ge \beta\ge \gamma >-\infty$ is assumed and 
$\sigma_{3}$ is a positive constant.
The above distribution 
is derived by assuming that each of the independent 
components of the deformation tensor has a Gaussian
distribution.
We should note that the probability for $\alpha$, $\beta$, and $\gamma$
to take values which are very close to each other 
is suppressed because of the 
factor $-(\alpha-\beta)(\beta-\gamma)(\gamma-\alpha)$.

It would be more comprehensive to rewrite the above in the following form:
 \begin{eqnarray}
&& w(\alpha,\beta,\gamma)d\alpha d\beta d\gamma \nonumber \\
&=& -\frac{27}{8\sqrt{5}\pi \sigma_{3}^{6}}\exp
\left[-\frac{1}{10\sigma_{3}^{2}}(\alpha+\beta+\gamma)^{2}
-\frac{1}{4\sigma_{3}^{2}}\{(\alpha-\beta)^{2}+(\beta-\gamma)^{2}+(\gamma-\alpha)^{2}\}
\right] \nonumber \\
&& \cdot (\alpha-\beta)(\beta-\gamma)(\gamma-\alpha)d\alpha d\beta d\gamma.
\end{eqnarray}
Introducing new variables $x$, $y$, and $z$ by 
\begin{equation}
 x=\frac{\alpha+\beta+\gamma}{3},~~
  y=\frac{(\alpha-\beta)-(\beta-\gamma)}{4}, ~~{\rm and}~~
  z=\frac{\alpha-\gamma}{2},
\label{eq:xyz}
\end{equation}
respectively, we find 
\begin{eqnarray}
&& w(\alpha,\beta,\gamma)d\alpha d\beta d\gamma =\tilde{w}(x,y,z)dxdydz \nonumber \\
&=&
-\frac{27}{\sqrt{5}\pi\sigma_{3}^{6}}
(2y-z)(2y+z)z\exp\left[-\frac{9}{10}\left(\frac{x}{\sigma_{3}}\right)^{2}
-2\left(\frac{y}{\sigma_{3}}\right)^{2}-\frac{3}{2}\left(\frac{z}{\sigma_{3}}\right)^{2}\right]dxdydz,
\label{eq:Doroshkevich_xyz}
\end{eqnarray}
where $-\infty<x<\infty$, $-\infty<y<\infty$, and $2|y|\le z<\infty$.
Using Eqs.~(\ref{eq:x2nGaussian_full}), (\ref{eq:xGaussian}), 
and (\ref{eq:x3Gaussian}),
we can explicitly show
\begin{equation}
 \int_{-\infty}^{\infty} d\alpha \int_{-\infty}^{\alpha} d\beta
  \int_{-\infty}^{\beta}d\gamma w(\alpha,\beta,\gamma)
=\int_{-\infty}^{\infty}dx \int_{-\infty}^{\infty}dy \int_{2|y|}^{\infty}dz
\tilde{w}(x,y,z)=1.
\end{equation}

\subsection{Integral expression and numerical integration}
\begin{figure}
 \begin{center}
  \includegraphics[width=0.7\textwidth]{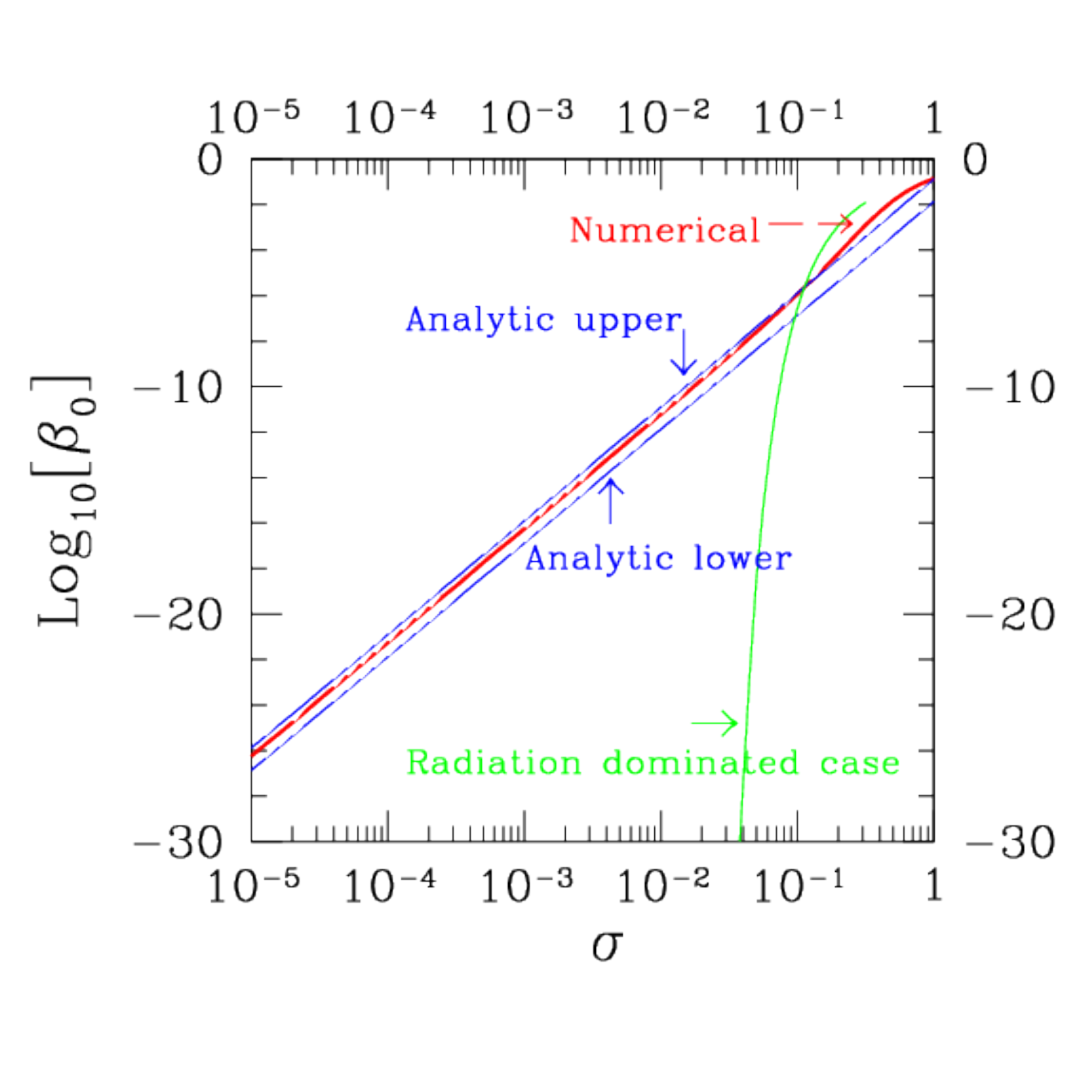}
\caption{\label{fg:pbh_rate} {
The production probability $\beta_{0}$ of primordial black holes in the
  matter-dominated phase of the Universe is plotted as a
  function of the density fluctuation $\sigma$ using the red
  line. 
  We find that for $\sigma\alt 0.01$, the numerical result agrees very well with a
  formula $\beta_{0}\simeq 0.0556\sigma^{5}$, which is semi-analytically
  obtained by imposing the constraint $\sigma\ll 1$. We also plot 
  lower and upper bounds which are analytically obtained for $\sigma\ll 1$
  using the blue lines. It should be noted that 
  if one would take into account an inhomogeneity 
  effect according to \citet{Khlopov:1980mg,Polnarev:1982}, one might have $\beta_{0}
  \sim \sigma^{13/2}$. The green line denotes a value of $\beta_{0}$ for the
  radiation-dominated phase discussed in
  Sec.~\ref{sec:radiation-dominated_phase}. The readers 
  should be cautioned that the value of $\beta_{0}$ 
  for the radiation domination (the green line) is obtained in spherical symmetry.}}
 \end{center}
\end{figure}

We define $\sigma$ as the standard deviation of $\delta_{L}(t_{i})$,
which is given by 
\begin{equation}
 \sigma^{2}=\overline{\delta_{L}^{2}(t_{i})}=\overline{(\alpha+\beta+\gamma)^{2}}\left(\frac{b}{a}\right)^{2}(t_{i})=5\sigma_{3}^{2},
\end{equation}
where we have used Eqs.~(\ref{eq:normalisation_b}) and 
(\ref{eq:Doroshkevich_distribution}).
The production probability of primordial black holes is
given by 
\begin{equation}
 \beta_{0}=\int_{0}^{\infty}d\alpha
  \int_{-\infty}^{\alpha}d\beta\int_{-\infty}^{\beta}d\gamma
  ~\theta(1-h(\alpha,\beta,\gamma))w(\alpha,\beta,\gamma),
\label{eq:full_order_expression}
\end{equation}
where we have used Eq.~(\ref{eq:criterion_3}).
Note that we still allow both a large deviation from 
spherical symmetry and a large amplitude for density perturbations. 
Note also that, in principle, 
one can add a condition $\alpha+\beta+\gamma>0$,
i.e., $\delta_{L}>0$ for black hole formation. Although this 
gives another factor 
$\theta(\alpha+\beta+\gamma)$ in the integrand in Eq.~(\ref{eq:full_order_expression}), it generally results in a very little change 
in the numerical value of $\beta_{0}$ and hence 
we proceed without this condition.
We have numerically implemented the triple integration to obtain 
$\beta_{0}$ and present the result in Fig.~\ref{fg:pbh_rate}. 
We can see that $\beta_{0}$ is a monotonically increasing function of $\sigma$.
For small $\sigma$, $\beta_{0}$ tends to be proportional 
to $\sigma^{5}$ and is best fit by 
\begin{equation}
\beta_{0}\simeq 0.056\sigma^{5}.
\label{eq:best_fit_curve} 
\end{equation}
As $\sigma$ increases beyond 0.01, $\beta_{0}$ becomes significantly 
larger than the above power law formula.

\subsection{Analytic estimate for $\sigma\ll 1$}
We can show that one of the three
integrals in the expression (\ref{eq:full_order_expression})
can be analytically implemented. 
If the constraint $\sigma\ll 1$ is imposed, 
we can obtain a semi-analytic estimate 
of $\beta_{0}$ by a single numerical integration. 
We here present the semi-analytic formula 
\begin{equation}
 \beta_{0}\simeq 0.05556 \sigma^{5}.
\label{eq:new_formula}
\end{equation} 
This confirmes the best-fit curve (\ref{eq:best_fit_curve}) for small $\sigma$. 
In the same framework, we can analytically derive lower and upper bounds
on $\beta_{0}$ so that  
\begin{equation}
 0.01338\sigma^{5}\alt \beta_{0}\alt 0.1280 \sigma^{5}.
\end{equation}
Since the derivation of 
the above results is rather technical, we postpone to 
describe it in the
Appendix~\ref{sec:semi-analytic_estimate}. 
Instead, we will give a comprehensive argument below, 
which provides us with a physical picture of 
black hole formation as well as 
the analytic lower and upper bounds on $\beta_{0}$.

Since $E(e)$ is a monotonically decreasing function of $e\in [0,1]$, 
Eq.~(\ref{eq:s_3}) implies 
\begin{equation}
 h(\alpha,\beta,\gamma)\ge  \frac{2}{\pi}\frac{\alpha-\gamma}{\alpha^{2}}.
\end{equation}
Therefore, the formation criterion $h\lesssim 1$ implies
\begin{equation}
 \alpha-\gamma \lesssim \frac{\pi}{2}\alpha^{2}.
\label{eq:upper_bound_nonsphericity}
\end{equation}
As we can see in Eq.~(\ref{eq:Doroshkevich_xyz}),
the probability that $x$, $y$, or $z$ takes a value 
much larger than $\sigma_{3}=\sigma/\sqrt{5}$ 
is exponentially suppressed.
Then, for $\sigma\ll 1$,  
it is most probable that 
$(\alpha+\beta+\gamma)=O(\sigma)$,
$(\alpha-\gamma)=O(\sigma)$, and $(\alpha-\beta)-(\beta-\gamma)=O(\sigma)$.
This means $\alpha=O(\sigma)$, $\beta=O(\sigma)$, and $\gamma=O(\sigma)$.
Therefore, Eq.~(\ref{eq:upper_bound_nonsphericity}) immediately
implies $\alpha-\gamma=O(\sigma^{2})$ and, hence,
$\alpha-\beta=(\sigma^{2})$ and $\beta-\gamma=O(\sigma^{2})$, 
since $\alpha\ge \beta\ge \gamma$.
Then, $\alpha$, $\beta$, and $\gamma$ are equal to each other to  
$O(\sigma)$. Therefore, the collapse is predominantly nearly 
spherically symmetric to $O(\sigma)$.
In this case, $\alpha>0$ implies $x>0$ to $O(\sigma)$.

Although the collapse is nearly spherical, it does not mean that 
the pancake is nearly circular but the size of the pancake is 
$\sim \sigma r_{f}$. 
The elliptic integral in Eq.~(\ref{eq:s_3}) 
for $h(\alpha,\beta,\gamma)$ makes it very difficult 
to directly integrate Eq. (\ref{eq:full_order_expression}). 
Here, we estimate the integral by using the inequality $E(0)\ge
E(e)\ge E(1)$ for $0\le e\le 1$. In the circular limit $e=0$, 
we can approximate $h$ as 
\begin{equation}
 h(\alpha,\beta,\gamma)\simeq \frac{\alpha-\gamma}{\alpha^{2}}.
\end{equation}
Therefore, the formation criterion $ h \lesssim 1$ implies
\begin{equation}
 \alpha-\gamma \lesssim \alpha^{2}.
\label{eq:simple}
\end{equation}
Then, in terms of $x$, $y$, and $z$, the domain for black hole formation
is given by 
\begin{equation}
 0<x,~~ -\frac{x^{2}}{4}<y<\frac{x^{2}}{4}, ~~
 2|y|<z<\frac{x^{2}}{2}.
\end{equation}
The probability of black hole formation can be calculated to 
\begin{equation}
\beta_{0}\simeq  \int_{0}^{\infty}dx
 \int_{-\frac{x^{2}}{4}}^{\frac{x^{2}}{4}} dy 
 \int_{2|y|}^{\frac{x ^{2}}{2}} dz\tilde{w}(x,y,z)\simeq \frac{7\cdot
  5^{3}}{2^{4}\cdot 3^{6}\sqrt{10\pi}}\sigma^{5}\simeq 0.01338 \sigma^{5},
\label{eq:analytic_formula_circular}
\end{equation}
where we have used the approximation $x^{2}\ll \sigma$
and Eq.~(\ref{eq:x2nGaussian_half}). 
In the eccentric limit $e=1$ of the pancake, 
$\beta_{0}$ can be similarly given by
\begin{equation}
\beta_{0}\simeq  \int_{0}^{\infty}dx
 \int_{-\frac{\pi}{2}\frac{x^{2}}{4}}^{\frac{\pi}{2}\frac{x^{2}}{4}} dy 
 \int_{2|y|}^{\frac{\pi}{2}\frac{x ^{2}}{2}} dz\tilde{w}(x,y,z)
\simeq \frac{7\cdot
  5^{3}}{2^{4}\cdot 3^{6}\sqrt{10\pi}}\left(\frac{\pi}{2}\right)^{5}
\sigma^{5}\simeq 0.1280 \sigma^{5}.
\label{eq:analytic_formula_eccentric}
\end{equation} 
Clearly, the circular 
and eccentric limits, 
(\ref{eq:analytic_formula_circular}) and 
(\ref{eq:analytic_formula_eccentric}),
correspond to the lower and upper
bounds, respectively. 
The semi-analytic formula (\ref{eq:new_formula}) implies that 
highly eccentric pancakes give a significant contribution to the probability of 
black hole formation. 

Note that 
there is a nonvanishing probability where $\alpha$, $\beta$, or $\gamma$
takes a value of the order greater than $\sigma$ and 
the resulting collapse is highly nonspherical.
The semi-analytic 
formula (\ref{eq:new_formula}) does not include this probability.
We should recall that 
the full expression (\ref{eq:full_order_expression}) includes 
all such cases. As seen in Fig.~\ref{fg:pbh_rate}, for
$\sigma\alt 0.01$, the semi-analytic formula 
(\ref{eq:new_formula}) agrees with the numerical result very well.
On the other hand, for $\sigma\agt 0.01$,
the numerical result is 
larger than the semi-analytic formula (\ref{eq:new_formula}).
This suggests that a highly nonspherical collapse results in 
black hole formation  
with a significant probability for $\sigma\agt 0.01$.

For $\sigma\ll 1$, as
the probability of black hole formation is dominated 
by near-spherical collapse, we can neglect 
gravitational radiation in the course of collapse 
because the energy gravitationally radiated away
from the mass is suppressed by a factor of $\sigma^{4}$. 
On the other hand, in 
the violent relaxation phase on and after the first caustic, 
gravitational radiation can be significantly large.
In the presence of a large velocity dispersion after the first caustic, 
it is highly nontrivial whether such gravitational radiation significantly
affects the picture of the virialization.
For the moment, we adopt the scenario where the mass once gets virialized 
through the violent relaxation and it may collapse to a black hole
in a time scale much larger than the Hubble time.
This scenario must be tested by numerical simulations 
and it is also very interesting to estimate gravitational radiation in
the nonspherical formation of primordial black holes and virialized 
objects.

\section{Discussion}

\subsection{Comparison with the Khlopov-Polnarev formula}
It is important to compare the present result with the previous 
result in the literature.
The Khlopov-Polnarev criterion for primordial black hole formation 
is given by~\citep{Khlopov:1980mg,Polnarev:1982,Khlopov:2008qy}
\begin{equation}
 \tilde{s}\lesssim \chi,
\end{equation}
where they put 
\begin{eqnarray}
 \tilde{s}:=
  \mbox{max}(|\alpha-\beta|,|\beta-\gamma|,|\gamma-\alpha|)~~
{\rm and}~~
 \chi:= \frac{r_{g}}{r_{f}}\simeq \delta_{L}(t_{i}).
\end{eqnarray}
From the above, they 
 derived the probability 
for primordial black hole formation as
\begin{eqnarray}
 \beta_{0}= \frac{27}{8\sqrt{5}\pi}\int_{0}^{\infty} 
d\alpha e^{-\frac{9}{10}\alpha^{2}}
\int_{\alpha-\alpha
\chi}^{\alpha}(\alpha-\beta)d\beta
\int_{\alpha-\alpha \chi}^{\beta}(\alpha-\gamma)(\beta-\gamma)d\gamma 
\simeq 0.02 \chi^{5}.
\label{eq:KP_formula}
\end{eqnarray}

Let us compare our new estimate with the Khlopov-Polnarev
one. We should first note that the new estimate is written in terms of
the density fluctuation $\sigma$, while the Khlopov-Polnarev
one is written in terms of $\chi$, which is dealt with 
as if it were 
a definitive value in \citet{Khlopov:1980mg,Polnarev:1982}. 
In spite of this critical conceptual difference, 
it is very curious that 
the new formula (\ref{eq:new_formula}) for $\sigma\ll 1$
agrees with the Khlopov-Polnarev one (\ref{eq:KP_formula})
only within a factor of 3 if we simply identify $\chi$ with $\sigma$. 
However, we would like to emphasize that 
the present analysis can deal with a large deviation 
from spherical symmetry and a large fluctuation $\sigma$ 
by adopting the hoop conjecture as the black hole formation criterion.

Here, we discuss an inhomogeneity effect on primordial black hole
formation. Khlopov and Polnarev \citep{Khlopov:1980mg,Polnarev:1982}
argued that if the central concentration within 
the overdense region is 
sufficiently high, a caustic, or a shell-focusing singularity in
modern terminology, occurs at the center of the mass before a 
black hole horizon is formed. They assumed that the equation of state 
then changes to that of 
radiation $p=\rho c^{2}/3$ due to the rise of the density 
in the central region and an arising pressure gradient would prevent the collapse from being 
a black hole. If the interaction between particles is not sufficiently strong, 
they assumed that particles escaping from the central region would also 
prevent black hole formation.
According to the above argument, they put 
another factor $\chi^{3/2}$ to the probability 
based on the analysis of the Lemaitre-Tolman-Bondi dust solution.
They finally obtained $\beta_{0}\simeq 0.02 \chi^{13/2}$.
This can be recast in the form $\beta_{0}\sim \sigma^{13/2}$ 
in terms of $\sigma$ in our formulation.
Although it is hard to generally exclude such a scenario, 
we only point out here that such an effect, if any, is highly 
dependent on the matter model.
It is also likely that, even if pressure arises in the central region, 
its gradient just slows down the collapse of the central 
region and eventually a black hole horizon forms as the surrounding 
layers fall down and accumulate on the central region.
In the current paper, we focus on the nonspherical effect
and simply neglect this model-dependent inhomogeneity effect.

\subsection{Comparison with the production rate in the radiation-dominated
 era} 
\label{sec:radiation-dominated_phase}

In the presence of relativistic pressure
$p=w\rho c^{2}$, where $w=1/3$ for radiation, 
the criterion of primordial black hole formation
is predominantly determined by the Jeans scale argument. 
Following \citet{Harada:2013epa,Harada:2015yda}, 
let us describe the criterion by using the density perturbation 
$\tilde{\delta}$ at horizon entry in the comoving slicing in the
long-wavelength limit, where $\tilde{\delta}$ is defined as
\begin{equation}
 \tilde{\delta}:=\lim_{\epsilon\to 0}\epsilon^{-2}\delta
\label{eq:tilde_delta}
\end{equation}
and $\epsilon:=ck/(aH)$. Note that $\delta$ is the density perturbation 
averaged over the comoving coordinate scale $k^{-1}$ and is 
proportional to $\epsilon^{2}$ for a fixed $k$.
By using the threshold value $\tilde{\delta}_{c}$, the criterion for black
hole formation is given by $\tilde{\delta}\agt \tilde{\delta}_{c}$,
where 
\begin{equation}
 \tilde{\delta}_{c}\simeq \frac{3(1+w)}{3w+5}\sin^{2}\left[\frac{\pi\sqrt{w}}{1+3w}\right],
\end{equation}
while the maximum value for $\tilde{\delta}$ is given by
\begin{equation}
 \tilde{\delta}_{\rm max}\simeq \frac{3(1+w)}{3w+5},
\end{equation}
if we neglect a relatively minor dependence on the density profile.
Although the above formula is 
based on spherical symmetry, nonspherical effects
are expected to be subdominant for the
relativistic pressure $w=O(1)$, because $\tilde{\delta}_{c}=O(1)$ 
and there is little time left for nonsphericity to sufficiently 
grow, even if it grows, until a black hole horizon forms.
If the density perturbation obeys a Gaussian
distribution, the production rate of primordial black holes is given
by~\citep{Carr:1975qj,Harada:2013epa}
\begin{equation}
 \beta_{0}\simeq
\int_{\tilde{\delta}_{c}}^{\tilde{\delta}_{\rm max}}d\tilde{\delta}
\frac{2}{\sqrt{2\pi}\sigma}\exp\left(-\frac{\tilde{\delta}^{2}}{2\sigma^{2}}\right)=
  {\rm
    erfc}\left(\frac{\tilde{\delta}_{c}}{\sqrt{2}\sigma}\right)-
  {\rm
    erfc}\left(\frac{\tilde{\delta}_{\rm max}}{\sqrt{2}\sigma}\right)\simeq 
   \sqrt{\frac{2}{\pi}}\frac{\sigma}{\tilde{\delta}_{c}}\exp\left(-\frac{\tilde{\delta}_{c}^{2}}{2\sigma^{2}}\right),
\label{eq:pbh_rate_radiation}
\end{equation}
where $\sigma^{2}:=\langle \tilde{\delta}^{2}\rangle $, 
the factor 2 comes from the Press-Schechter argument,
${\rm erfc}(x)$ is the complementary error function 
and the rightest expression is valid only for $(\tilde{\delta}_{\rm
max}-\tilde{\delta}_{c})/\sigma \gg 1$ and 
$\tilde{\delta}_{c}/ \sigma \gg 1 $.
We plot the second left expression in 
Eq.~(\ref{eq:pbh_rate_radiation}) 
as a function of $\sigma$ in the radiation-dominated era 
in Fig.~\ref{fg:pbh_rate} with the green line, 
where $w=1/3$, $\tilde{\delta}_{c}\simeq 0.4135$ and
$\tilde{\delta}_{\rm max}=2/3$ are chosen.

We can see that the production probability in the radiation-dominated era 
is smaller than that in the matter-dominated era for $\sigma\alt 0.05$.
This can be understood as an effect of 
the relativistic pressure which suppresses the 
collapse of the overdense region to a black hole.
On the other hand, for $\sigma\agt 0.05 $,
the graph seems to show that the production probability 
in the matter-dominated era 
is smaller than that in the radiation-dominated era.
Although we should be cautioned that nonspherical effects 
in the radiation-dominated era are simply neglected in 
Eq.~(\ref{eq:pbh_rate_radiation}), 
we can conclude that the production of primordial black 
holes in the matter-dominated era can be suppressed by the nonspherical
effect so strongly that the production rate 
can be as small as or smaller than that in the radiation-dominated era
for $\sigma\agt 0.05$. To clarify this issue, it will be important to 
investigate nonspherical effects on primordial black hole 
formation in the presence of relativistic pressure.
It should also be noted that as indicated by \citet{Kopp:2010sh}, 
there is a serious problem in the conventional 
assumption of the Gaussian distribution for 
the density perturbation because of the presence of its maximum value
and its double-valued nature, in particular for large amplitude of fluctuation. 
It is also important to investigate this issue further.

\subsection{Density perturbation in terms of primordial curvature 
perturbation} 
\label{sec:perturbation}

Here, let us first briefly 
introduce the relativistic cosmological perturbation theory
based on \citet{Lyth:2004gb} and \citet{Harada:2015yda}.
In the long-wavelength limit $\epsilon\to 0$, 
the density perturbation $\delta$
can be written in the comoving slicing as~\citep{Harada:2015yda}
\begin{equation}
 \delta=-\frac{4(1+w)}{3w+5}\frac{c^{2}}{a^{2}H^{2}}\frac{\Delta\psi}{\psi^{5}}
\label{eq:delta_long_wavelength_limit_psi}
\end{equation}
for the equation of state $p=w\rho c^{2}$, 
where $\Delta$ is the Laplacian in the
flat 3-space, the curvature variable $\psi$ is defined 
so that the spatial metric $\gamma_{ij}$ is written in a form 
$\gamma_{ij}=\psi^4 a^2 \tilde{\gamma}_{ij}$ by introducing
the conformal spatial metric $\tilde{\gamma}_{ij}$ whose determinant is equal to
that of the metric in the flat 3-space, 
and the decaying mode is neglected. 
In the long-wavelength limit, 
$\psi$ takes an identical value, 
whether it is described 
in the comoving slicing, uniform-density slicing or 
constant-mean-curvature slicing, and is also 
conserved in the above slicings for adiabatic perturbation.

The curvature perturbation $\zeta$ can be defined by 
\begin{equation}
 e^{-2\zeta}=\psi^{4}
\end{equation}
under the uniform-density slicing condition~\citep{Lyth:2004gb}. 
This $\zeta$ is often used for calculating primordial
cosmological perturbations generated in inflation.
In terms of $\zeta$, 
Eq.~(\ref{eq:delta_long_wavelength_limit_psi}) is rewritten in the form
\begin{equation}
 \delta=-\frac{4(1+w)}{3w+5}\frac{c^{2}}{a^{2}H^{2}}e^{(5/2)\zeta}\Delta e^{-\zeta/2}.
\label{eq:delta_long_wavelength_limit_zeta}
\end{equation}
We can calculate $\tilde{\delta}_{k}$
given by Eq.~(\ref{eq:tilde_delta}) and
its variance by $\sigma_{k}^{2}=\langle \tilde{\delta}_{k}^{2}\rangle$, 
where the comoving wave number $k$ is explicitly indicated.
Note that the above 
argument so far does not require $\zeta$ to be small.
If we additionally linearize 
Eq.~(\ref{eq:delta_long_wavelength_limit_zeta}) with respect to $\zeta$, 
we find 
\begin{equation}
 \tilde{\delta}_{k} =2
\left.\left(\frac{1+w}{3w+5}\right)\zeta_{k}\right|_{k=aH(t_{i})}
\end{equation}
up to a phase factor and, hence, 
\begin{equation}
 \sigma_{k}^{2} =4\left(\frac{1+w}{3w+5}\right)^{2}
\langle \zeta_{k}^{2}\rangle_{k=aH(t_{i})}.
\label{eq:sigma2_zeta_k}
\end{equation}

Noting that the Zel'dovich approximation in the matter-dominated phase 
correctly describes the density
perturbation in the linear regime in the Newtonian gravity and this 
coincides with that in the relativistic perturbation theory in 
the comoving slicing both at subhorizon and superhorizon 
scales~\citep{Peebles:1980,Hwang:2012},
we can adopt an identification
\begin{equation}
 \delta_{L}(t_{i})=\delta (t_{i}).
\end{equation}
Therefore, we can identify the 
density perturbation in the Newtonian cosmology with that in the 
comoving slicing in the general relativistic perturbation theory 
in the linear regime. In fact, in the Newtonian gauge with the 
Newtonian potential $\phi$, we can find $\zeta=(5/3)\phi/c^{2}$ 
in the matter-dominated phase of the Universe in the linear regime. 
By linearizing Eq.~(\ref{eq:delta_long_wavelength_limit_zeta}), 
we can recover the Poisson equation both in subhorizon and superhorizon 
scales as
\begin{equation}
 \frac{1}{a^{2}}\Delta \phi=4\pi G \bar{\rho} \delta.
\end{equation}

In summary, 
in the radiation-dominated phase, 
Eq.~(\ref{eq:sigma2_zeta_k}) with $w=1/3$ reduces to
\begin{equation}
 \sigma_{k}^{2}=\frac{16}{81}\langle \zeta_{k}^{2}\rangle_{k=aH(t_{i})},
\end{equation}
where $\sigma_{k}$ is the standard deviation of the 
density perturbations at horizon entry in the linear regime
in the comoving slicing in the general relativistic perturbation theory.
In the matter-dominated phase, 
Eq.~(\ref{eq:sigma2_zeta_k}) with $w=0$ reduces to
\begin{equation}
 \sigma_{k}^{2}=\frac{4}{25}\langle \zeta_{k}^{2}\rangle_{k=aH(t_{i})}
\end{equation}
and $\sigma_{k}$ can be identified with the standard deviation of
the density perturbations at horizon entry in the linear regime 
in the Newtonian cosmology.

\section{Conclusion}
We have studied primordial black hole formation in the matter-dominated
era of the Universe. In this epoch, in the absence of relativistic 
pressure, nonspherical effects play a crucial role and 
gravitational collapse does not necessarily lead to black hole
formation. We have applied the Zel'dovich approximation to the 
nonlinear evolution of density perturbations in an expanding universe,
Thorne's hoop conjecture for the formation of a black hole horizon,
and Doroshkevich's probability distribution for the eigenvalues of the 
deformation tensor.
We have succeeded in obtaining an integral expression for the 
probability of black hole formation, which allows for a
large fluctuation of density perturbations
and a large deviation from spherical symmetry.
We have plotted the result of the numerical integration for 
the production probability as a function of the density fluctuation.
Moreover, we have obtained a compact semi-analytic
formula for a small fluctuation, which is comparable with 
Khlopov and Polnarev's formula with some nontrivial 
identification of parameters.
We have also analytically obtained lower and upper bounds for 
a small fluctuation.
This implies that the current analysis has essentially 
refined Khlopov and Polnarev's heuristic 
argument and generalized it to 
a large deviation from spherical symmetry 
and a large fluctuation of density perturbations.
Both the integral expression and semi-analytic
formula are applicable for the 
estimate of abundance of primordial black holes in the matter-dominated
era of the Universe, such as the first-order phase transition, 
the ending phase of inflation before reheating and the late-time
matter-dominated era following the matter-radiation equality.
We have compared the new estimate of the production rate 
with that in the radiation-dominated phase of the Universe
and found that the matter dominance strongly enhances
primordial black hole formation for small density fluctuation, 
while it does not for larger density fluctuation.
We have also presented a formula which gives the initial density
fluctuation in terms of primordial curvature perturbation.

\acknowledgments
The authors would like to thank W.~East, I.~Musco, T.~Nakama, and S.~Yokoyama 
for fruitful discussion. This work was supported by JSPS
KAKENHI Grant Numbers JP26400282 (T.H.), JP16K17688, JP16H01097 (C.Y.), 
JP26247041, JP15H05889, JP16H0877 (K.K.), and
JP25400265 (K.N.).
\appendix

\section{Formulas of Gaussian integrals}
\label{sec:Gaussian_integral}

We present the following well-known
formulas for Gaussian integrals: 
\begin{eqnarray}
 \int_{-\infty}^{\infty}x^{2n}e^{-\alpha x^{2}}dx&=&
\frac{(2n-1)!!}{(2\alpha)^{n}}\sqrt{\frac{\pi}{\alpha}}, 
\label{eq:x2nGaussian_full}
\\
\int_{0}^{\infty}x^{2n}e^{-\alpha
 x^{2}}dx&=&\frac{1}{2}
\frac{(2n-1)!!}{(2\alpha)^{n}}\sqrt{\frac{\pi}{\alpha}}, 
\label{eq:x2nGaussian_half}
\\
\int x e^{-\alpha  x^{2}}dx&=&-\frac{1}{2\alpha}e^{-\alpha x^{2}}, 
\label{eq:xGaussian}
\\
\int x^{3}e^{-\alpha  x^{2}}dx&=&
-\frac{1}{2\alpha^{2}}(1+\alpha x^{2})e^{-\alpha x^{2}}, 
\label{eq:x3Gaussian} 
\\
\int x^{5}e^{-\alpha  x^{2}}dx&=&
-\frac{1}{\alpha^{3}}(2+2\alpha x^{2}+\alpha^{2} x^{4})e^{-\alpha
x^{2}},
\label{eq:x5Gaussian}
\end{eqnarray}
where $\alpha>0$, $n$ is a nonnegative integer and we have 
put $0!!=(-1)!!=1$.

\section{Estimate of the integral in $\beta_{0}$}
\label{sec:semi-analytic_estimate}

To estimate $\beta_{0}$, we further change the variables from $(x,y,z)$
given by Eq.~(\ref{eq:xyz}) to $(t,u,z)$ defined as
\begin{equation}
 t=\frac{x}{z},~~  u=\frac{y}{z}, ~~z=z,
\end{equation}
where the domain is given by 
$-\infty < t < \infty$, $-1/2 < u < 1/2$, and $0<z<\infty$. 
The probability distribution is rewritten in the form 
\begin{eqnarray}
\bar {w}(t,u,z)dtdudz  =
-\frac{27}{\sqrt{5}\pi\sigma_{3}^{6}}
(2u-1)(2u+1)z^5\exp\left[-A(t,u)z^2\right]dtdudz, 
\label{eq:Doroshkevich_tuz}
\end{eqnarray}
where 
\begin{equation}
A(t,u):=\frac{9}{10}\left(\frac{t}{\sigma_{3}}\right)^{2}
+2\left(\frac{u}{\sigma_{3}}\right)^{2}+\frac{3}{2}\left(\frac{1}{\sigma_{3}}\right)^2. 
\end{equation}
Since 
\begin{equation}
h=\frac{4}{\pi z}
\left(t+\frac{2}{3}u+1\right)^{-2}E\left(\sqrt{1-\left(u+\frac{1}{2}\right)^2}\right), 
\end{equation}
the criterion $h<1$ can be written in the following form:
\begin{equation}
z>z_*(t,u):=\frac{4}{\pi}\left(t+\frac{2}{3}u+1\right)^{-2}E\left(\sqrt{1-\left(u+\frac{1}{2}\right)^2}\right). 
\end{equation}
We also find that $\alpha>0$ implies 
$
t>-1-(2/3)u.
$
Therefore,
$\beta_{0}$ can be calculated as follows:
\begin{eqnarray}
\beta_0&=&-\frac{27}{\sqrt{5}\pi\sigma_3^6}\int^{1/2}_{-1/2} 
du (2u-1)(2u+1) 
\int^\infty_{-1-\frac{2}{3}u} dt \int^\infty_{z_*} dz 
z^5 e^{-A z^2} \cr
&=&-\frac{27}{2\sqrt{5}\pi\sigma_3^6}\int^{1/2}_{-1/2} 
du (2u-1)(2u+1) \int^\infty_{-1-\frac{2}{3}u} dt 
\frac{2+2Az_*^{2}+A^2z_*^4}{A^3}e^{-Az_*^2}, 
\label{eq:beta_0_double_integral}
\end{eqnarray}
where we have used Eq.~(\ref{eq:x5Gaussian}) 
and omitted the arguments
$(t,u)$ of $A$ and $z_{*}$. Thus, the integration with respect to $z$ is
done.

For $\sigma\ll 1$, we can find that the dominant contribution 
in the $t$-integral comes from 
the region $t\agt \sigma^{-1}$ because the contribution from the outside
of this region is exponentially suppressed. 
For this region, since $t\gg 1$, we find 
\begin{equation}
A \simeq  \frac{9}{10}\left(\frac{t}{\sigma_{3}}\right)^{2},~~
z_* \simeq \frac{4}{\pi}\frac{E}{t^{2}},~~{\rm and}~~
A z_*^2\simeq \frac{72}{5\pi^2 \sigma_3^2}\frac{E^2}{t^{2}}, 
\end{equation}
where we have omitted the argument of $E$.
Since the contribution to the integral with respect to $t$ in
Eq.~(\ref{eq:beta_0_double_integral}) 
from the interval $[-1-(2/3)u,0]$ is negligible,  
changing the variable from $t$ to $w=1/t$, we obtain 
\begin{eqnarray}
\beta_0&\simeq &
-\frac{8\sqrt{5}}{27\pi^5\sigma_3^4}\int^{1/2}_{-1/2} 
du (2u-1)(2u+1)\int^\infty_0 dw w^{4}  \nonumber \\ 
&& 
\times \left(25\pi^4\sigma_3^4+360\pi^2\sigma_3^2w^2E^2+2592w^4E^4\right)\exp\left(-\frac{72w^2E^2}{5\pi^2\sigma_3^2}\right)
\nonumber \\
&=&\frac{7\cdot 5^{5} \pi^{9/2}}{2^{10}\cdot
3^{5}\sqrt{2}}\sigma_3^5\int^{1/2}_{-1/2} du (1-2u)(1+2u)E^{-5} 
= \frac{7\cdot 5^{3}\pi^{9/2}}{2^{9}\cdot
3^{6}\sqrt{10}}\bar{E}^{-5}\sigma^{5},
\label{eq:semi-analytic_estimate}
\end{eqnarray}
where we have used $\sigma=\sqrt{5}\sigma_{3}$ and 
Eq.~(\ref{eq:x2nGaussian_half}) and 
defined $\bar{E}$ as
\begin{equation}
 \bar{E}^{-5}:=\frac{3}{2}\int^{1/2}_{-1/2} du (1-2u)(1+2u)E^{-5}\left(\sqrt{1-\left(u+\frac{1}{2}\right)^2}\right).
\end{equation}
The direct numerical integration with respect to $u$ gives
Eq.~(\ref{eq:new_formula}) with $\bar{E}\simeq 1.182$, 
while it is clear that if we replace $\bar{E}$ in the last expression in
Eq.~(\ref{eq:semi-analytic_estimate}) with the circular limit value 
$\pi/2$ and eccentric limit value $1$, we obtain the same 
lower and upper bounds as are given by
Eqs. (\ref{eq:analytic_formula_circular}) 
and (\ref{eq:analytic_formula_eccentric}), respectively.

\end{document}